\documentclass[conference,anonymous]{IEEEtran}
\usepackage{textcomp} 
\usepackage{nicefrac}
\usepackage[utf8]{inputenc}
\usepackage[english]{babel}
\PassOptionsToPackage{bookmarks=false}{hyperref}
\usepackage{multirow, multicol, booktabs, tabulary, tabu, longtable, array, varwidth}
\usepackage{placeins, lipsum}
\usepackage[flushleft]{threeparttable}
\usepackage{tabularx}
\usepackage{amsmath}
\usepackage[hyphens]{url}
\usepackage[hidelinks]{hyperref}
\hypersetup{breaklinks=true}
\usepackage{cite}
\usepackage[hyphens]{url}
\usepackage[labelfont=bf]{caption}
\usepackage{cite}
\usepackage{hyperref}
\usepackage[dvipsnames, table]{xcolor}
\hypersetup{
    linkcolor  = violet!85!black,
    citecolor  = magenta!85!black,
    urlcolor   = blue!85!black,
    colorlinks = true,
    breaklinks = true
}

\usepackage{textcomp}
\usepackage[shortcuts,acronym]{glossaries}
\usepackage{soul, footnote, xargs}

\usepackage{amsmath}
\usepackage[utf8]{inputenc}
\usepackage{enumerate}

\usepackage[inline]{enumitem}

\usepackage{amsmath, amsfonts, amssymb, amsthm, nicefrac}
\usepackage{siunitx}
\theoremstyle{definition}
\newtheorem{definition}{Definition}

\usepackage{listings}

\usepackage[scaled=0.82]{beramono}

\usepackage[]{cleveref} %
\crefname{appsec}{Appendix}{Appendices}
\crefformat{section}{\S#2#1#3}
\crefformat{subsection}{\S#2#1#3}
\crefformat{subsubsection}{\S#2#1#3}
\crefformat{equation}{(#2#1#3)}
\crefrangeformat{equation}{(#3#1#4--#5#2#6)}
\crefmultiformat{equation}{(#2#1#3)}{ and~(#2#1#3)}{, (#2#1#3)}{ and~(#2#1#3)}
\crefformat{figure}{Fig.~#2#1#3}
\crefrangeformat{figure}{Figs. #3#1#4--#5#2#6}
\crefmultiformat{figure}{Figs.~#2#1#3}{ and~#2#1#3}{, #2#1#3}{ and~#2#1#3}
\crefname{algocf}{Alg.}{Algs.}
\Crefname{algocf}{Algorithm}{Algorithms}
\crefformat{table}{Table~#2#1#3}
\crefrangeformat{table}{Tables~#3#1#4--#5#2#6}
\crefmultiformat{table}{Tables~#2#1#3}{ and~#2#1#3}{, #2#1#3}{ and~#2#1#3}

\usepackage{textcomp}
\usepackage[shortcuts,acronym]{glossaries}
\usepackage{soul, footnote, xargs}
\usepackage[colorinlistoftodos,prependcaption,textsize=tiny]{todonotes}
\usepackage{balance}

\usepackage{graphicx}
\usepackage{subcaption}
\usepackage{tikz, epstopdf, stfloats, bbding, capt-of}
\usepackage{algorithmic}
\usepackage[ruled, linesnumbered]{algorithm2e}

\SetCommentSty{mycommfont}
\usepackage{microtype}
\usepackage[T1]{fontenc}

\def\baselinestretch{0.915}

\makeatletter
\newcommand\subparagraph{%
    \@startsection{subparagraph}{5}
    {\parindent}
    {3.25ex \@plus 1ex \@minus .2ex}
    {-1em}
    {\normalfont\normalsize\bfseries}}
\makeatother
\usepackage[compact]{titlesec}
\let\subparagraph\relax %
\titlespacing\section{0pt}{6pt plus 4pt minus 2pt}{2pt plus 2pt minus 2pt}
\titlespacing{\subsection}{0pt}{4pt plus 2pt minus 1pt}{2pt plus 1pt minus 1pt}
\titlespacing{\subsubsection}{0pt}{4pt plus 2pt minus 1pt}{2pt plus 1pt minus 1pt}
\usepackage{etoolbox}
\makeatletter
\patchcmd{\ttlh@hang}{\parindent\z@}{\parindent\z@\leavevmode}{}{}
\patchcmd{\ttlh@hang}{\noindent}{}{}{}
\makeatother

\usepackage{soul}
\sethlcolor{black}
\makeatletter
\newif\if@blind
\@blindtrue %
\if@blind \sethlcolor{black}\else
   
\fi
\newcommand{\spacefigure}{\vspace{0mm}}

\newcommand*\circled[1]{\tikz[baseline=(char.base)]{
        \node[shape=circle,draw,inner sep=1.2pt, Black, fill=Black] (char)
               {\color{white}\scriptsize\textbf{#1}};}%
        }

\newcommand{\BLUE}[1]{{\color{black} #1}}

\newcommand{\sysname}{{Netscope}\xspace}

\def\BibTeX{{\rm B\kern-.05em{\sc i\kern-.025em b}\kern-.08em
    T\kern-.1667em\lower.7ex\hbox{E}\kern-.125emX}}
\begin{document}

\title{
Application-aware Congestion Mitigation for High-Performance Computing Systems}
\author{
Archit Patke$^1$, 
Saurabh Jha$^1$, 
Haoran Qiu$^1$, 
Jim Brandt$^2$, 
Ann Gentile$^2$, \\
Joe Greenseid$^3$, 
Zbigniew T. Kalbarczyk$^1$, 
Ravishankar K. Iyer$^1$\\
$^1$\textit{University of Illinois at Urbana-Champaign}, 	$^2$\textit{Sandia National Laboratories}, $^3$\textit{Cray/HPE Inc. }

}

\maketitle
\begin{abstract}
High-performance computing (HPC) systems frequently experience congestion leading to significant application performance variation. However, the impact of congestion on application runtime differs from application to application depending on their network characteristics (such as bandwidth and latency requirements).
We leverage this insight to develop \sysname, an automated ML-driven framework that considers those network characteristics to dynamically mitigate congestion.
We evaluate \sysname on four Cray Aries systems, including a production supercomputer on real scientific applications. \sysname has a lower training costs and accurately estimates the impact of congestion on application runtime with a correlation between 0.7 and 0.9 for common scientific applications. Moreover, we find that \sysname reduces tail runtime variability by up to 14.9\texttimes~while improving median system utility by 12\%.

\end{abstract}

\section{Introduction}
\label{s:introduction}
Modern high-performance computing (HPC) systems concurrently execute multiple distributed applications that contend for the high-speed network leading to congestion. 
To limit contention in the network, several control (CC) mechanisms that limit application traffic injection into the network have been proposed~\cite{jiang2015network,luo2012congestion,escudero2014efficient,alizadeh2010data,mittal2015timely,kumar2020swift}.
However, application runtime variability (i.e., variable job completion times for the exact same input parameters) and suboptimal system utility (i.e., increased node hours to complete workloads) continue to be a problem in production systems~\cite{jha2020measuring,bhatele2020case}.
For example, runtime variability of up to 100\% has been observed in a 512 node MILC application~\cite{milc}, which projects to several million node hours of wasted compute time (based on~\cite{nersc_workload}).

\textbf{Insights and Challenges.} 
A significant limitation of prior work is its \textit{sole} reliance on a partial network view (e.g., message round trip time on application placement-defined paths) to perform congestion mitigation without considering the end-to-end adverse effects of congestion on application runtime.
Our network measurements show that end-to-end effects of congestion can significantly degrade the performance of some applications while having minimal impact on others (described in \cref{s:motivation}).
By not considering variability in end-to-end effects of congestion on application runtime, current CC mechanisms incorrectly penalize (i.e., throttle traffic of) applications  that are more susceptible to effects of congestion (described in \cref{s:motivation}).
Thus, congestion mitigation must include end-to-end effects as an important feature in modeling and mitigating an application's runtime variability.
However, it is challenging to infer end-to-end adverse effects of congestion on application runtime dynamically as it requires analysis and modeling of a high-dimensional dataset consisting of network telemetry data (e.g., network performance counters), network topology, and application node placement.

\textbf{Our Approach.}
This paper proposes a machine learning (ML) method that estimates end-to-end effect of congestion by combining a probabilistic model with domain insights and regression, which is trained using both production and synthetically-congested application executions (that includes network telemetry, topology, performance counters, and node placement).
Inference on this model drives an application-aware congestion mitigation mechanism that:
\begin{enumerate*}[label=(\roman*)]
    \item rate limits (i.e., throttles) application traffic that will not be adversely affected by reduced bandwidth, and
    \item avoids traffic rate limiting of applications that are more sensitive to congestion.
\end{enumerate*}

We implement this machine learning-based application-aware congestion mitigation framework as \sysname. \sysname has the following two components: an \emph{application behavior inference engine} and a \emph{congestion mitigation mechanism}.
\par \textbf{Application Behavior Inference engine.}
The application behavior inference engine estimates the end-to-end effect of congestion on application runtime using an ML model.
Typically, training this ML model requires several training runs that can be prohibitively expensive to obtain in production settings. 
For example, AutoML~\cite{he2019automl} and other black-box models~\cite{bhatele2020case} would require thousands of application runs as training input. 
We address the challenge of reducing application runs required for training by:
\begin{enumerate*}[label=(\roman*)]
\item \emph{distinct but inter-linked} system-specific and application-specific congestion impact estimation models. 
The system model estimates the impact of congestion on message delivery using network telemetry data. 
The application model uses inference from the system model to estimate the impact of congestion on application runtime.
This separation allows us to train the system model for a given HPC system using a specialized \texttt{pingpong} application (described in \cref{s:results}). 
This trained system model is then leveraged to optimize the training of the application model, thereby requiring significantly fewer additional application runs.
\item \emph{synthetically-creating congestion} to generate adverse congestion scenarios to accelerate model training. 
These scenarios allow us to explore the diverse possibilities of congestion quickly compared to production settings.
\end{enumerate*}

\par \textbf{Congestion Mitigation.}
The congestion mitigation mechanism uses a dynamic traffic rate limiter based on an Additive Increase Multiplicative Decrease (AIMD)\footnote{An AIMD control loop exponentially throttles application traffic (multiplicative decrease) when congestion is sensed in the network, and linearly increases traffic (additive increase) in the absence of detected congestion.}~\cite{chiu1989analysis} control loop whose \emph{parameters are inferred using our application behavior inference engine}. 
\sysname enhances the state-of-the-art AIMD mechanism by making the multiplicative decrease step in the AIMD control loop dependent on the end-to-end effects of congestion on application runtime. 
The enhanced AIMD significantly reduces application runtime variability and increases overall system utility, especially in the tail.

\textbf{Results.}
We implement and evaluate \sysname on four Cray Aries systems~\cite{alverson2012cray} including a production supercomputer across real applications; which includes MILC~\cite{milc}, Quantum Espresso~\cite{quantum_benchmarking}, AMG~\cite{amg}, and LAMMPS~\cite{lammps}. 
We also plan to release our code post paper acceptance.
Our evaluation shows:
\begin{enumerate}[label=(\alph*)]
\item \emph{The congestion mitigation mechanism significantly reduces runtime variability and improves system utility.}
In our evaluation, tail application runtime variability with \sysname is 14.9\texttimes~lower than with Cray static rate control and 11.2\texttimes~lower than with DCQCN rate control~\cite{zhu2015congestion} (a CC mechanism for RDMA networks). Moreover, median node hours used with \sysname are 8\% and 12\% lower than with Cray static rate control and DCQCN, respectively.

\item \emph{The application behavior inference engine accurately estimates the impact of congestion on application runtime.}
The application behavior inference engine can estimate application runtime with a correlation between 0.7 and 0.9 for common scientific applications under synthetically-generated congestion and natural system congestion found in production settings.

\item \emph{The application behavior inference engine has a low training cost.}
The engine can adapt to a new application within thirty runs in production environments and five runs with synthetically generated congestion.
\end{enumerate}

\section{Motivation}
\label{s:motivation}
Applications running in HPC systems~\cite{nersc_workload} have diverse communication characteristics in terms of varying message sizes, interplay between communication primitives, and application logic.
Those applications can be broadly categorized into \emph{delay sensitive} and \emph{bandwidth intensive} applications.
\emph{Delay sensitive} applications rely on low link buffer occupancy to achieve minimal network latency and low application runtime (i.e. increase in network latency increases application runtime significantly).
Examples include  MILC~\cite{milc} and Quantum Espresso~\cite{quantum_benchmarking,coral_2}.
We provide a definition and quantification of delay sensitivity in~\cref{s:framework}.
\emph{Bandwidth intensive} applications rely on large message sizes and collective communication primitives~\cite{pjevsivac2007performance} that exhausts available network bandwidth.
Examples include spectral codes~\cite{doi:10.1146/annurev.fluid.30.1.539} that use all-to-all traffic patterns.
GPCNeT~\cite{chunduri2019gpcnet} is a benchmark that mimics such bandwidth intensive communication patterns (i.e., all-to-all and N-to-1) found in HPC applications. 
Moreover, there exists an entire spectrum of applications that can be found at varying levels of delay sensitivity and bandwidth intensity. 
These applications contend with each other for limited network resources, and failure to efficiently multiplex network resources leads to runtime variability and suboptimal system utility. 
To address those problems, we leverage the following insights to design an application-aware congestion mitigation framework.

\begin{figure*}[ht]
	\begin{subfigure}[c]{0.325\textwidth}
    \includegraphics{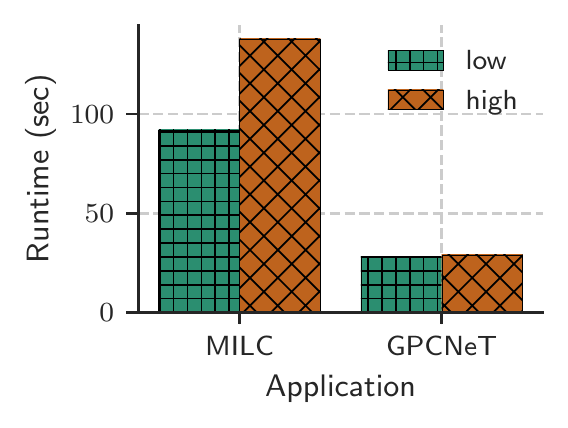}
    \caption{Variable impact of congestion on MILC and GPCNeT runtime as captured by low and high network stall counters}
    \label{fig:network_counters}
    \end{subfigure}
	\begin{subfigure}[c]{0.325\textwidth}
    \includegraphics{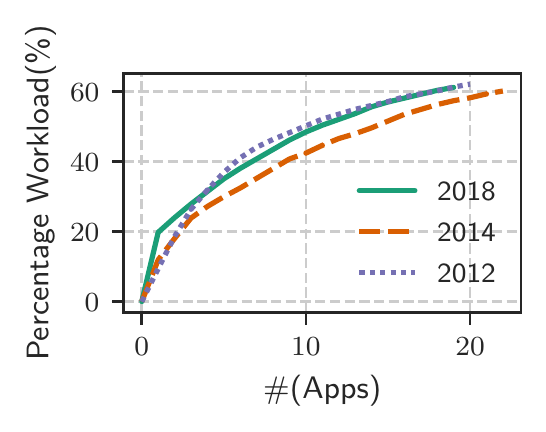}
    \caption{Workload composition on three generations of NERSC production supercomputers}
    \label{fig:workload_composition}
    \end{subfigure}
	\begin{subfigure}[c]{0.325\textwidth}
    \includegraphics{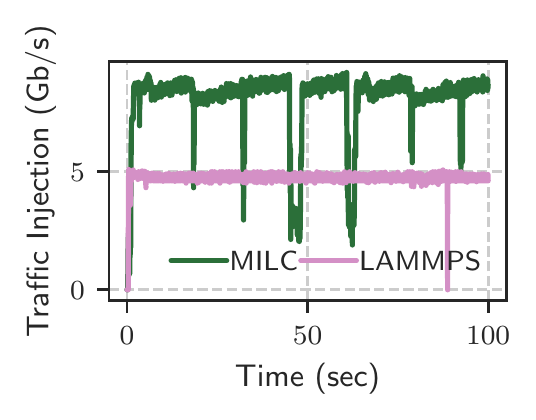}
    \caption{Variation in cumulative per node traffic injection across the application's lifespan for MILC and LAMMPS}
    \label{fig:traffic_variation}
    \end{subfigure}
\caption{Characterizing high-performance computing workloads}
\end{figure*}

\textbf{Insight 1: Runtime variability can be reduced by augmenting the CC mechanism with application delay sensitivity.}
Congestion signals used in rate control mechanisms such as ECN~\cite{ramakrishnan2001addition}, QCN~\cite{pan2007qcn}, and network delay~\cite{mittal2015timely,kumar2020swift} do not address the source of congestion directly.
Consequently, during the operation of the CC mechanism, the non-congestion-contributing applications are penalized along with the culprit that is responsible for the congestion. 
Moreover, if the application that was not responsible for congestion is also delay sensitive, it's performance degradation due to congestion mitigation can be severe. 
Because of this incorrect penalization, use of CC has been disabled on Sierra and Summit~\cite{chunduri2019gpcnet} systems (ranked second and third among the ``TOP500'' supercomputers~\cite{top_500}).
To resolve the incorrect penalization problem, we created an \emph{application-aware} rate control mechanism (described in \cref{s:mitigation}) that penalizes applications based on their delay-sensitivity, and tested it on a Cray Aries~\cite{alverson2012cray} HPC system (16 switches and 54 nodes) against two different algorithms (Cray static rate control~\cite{alverson2012cray} and DCQCN~\cite{zhu2015congestion}).
Our test workload consists of a delay sensitive application (MILC, isolated runtime 26 sec) and a bandwidth intensive application (GPCNeT, isolated runtime 75 sec) that execute simultaneously. 
Overall (see Table~\ref{tab:congestion_comparison}),  we find that the application-aware mechanism ensures low runtime variability for both applications. 
Our application-aware rate control throttled (i.e., penalized) MILC traffic only 20\% of what it throttled GPCNeT.

\begin{table}[ht]
\spacefigure
\centering
{%
\begin{tabular}{@{}cccc@{}}
\toprule
& Cray   & DCQCN  & \begin{tabular}[c]{@{}c@{}}Application-Aware\end{tabular} \\ \midrule
\begin{tabular}[c]{@{}c@{}}MILC (Baseline: 26 sec)\end{tabular}   & 115\% & 58\% & 7\% \\
\begin{tabular}[c]{@{}c@{}}GPCNeT (Baseline: 75 sec)\end{tabular} & 0\% & 2.7\% & 2.7\% \\
\bottomrule
\end{tabular}%
}
\caption{Measured percentage increase in application runtime with respect to baseline for three CC mechanisms.} 
\label{tab:congestion_comparison}
\vspace{-4mm}
\end{table}

\textbf{Insight 2: Network performance counter\footnote{Network performance counters such as link stall times and traffic can be collected at switches via vendor-provided mechanisms.} telemetry can be used to estimate application delay sensitivity.}
Network performance counters are widely collected on production systems~\cite{agelastos2014lightweight} and can be used to characterize network  behaviour. \cref{fig:network_counters} shows the runtime for MILC and GPCNeT under low (0.25\% percent time stalled\footnote{Percent Time Stalled is a proxy for measuring congestion in credit-based flow control networks that we describe in \cref{s:model}}) and high  (7.5\% percent time stalled) congestion, as measured from network stall counters. 
The figure shows that response to congestion in the system (that is captured through network counters) varies.
For instance, MILC has a 60\% increase in runtime, while GPCNeT's runtime marginally increases by 2.6\% for the same difference in measured congestion. 
We can leverage this variability in behavior as measured from network performance counters to enable estimation of delay sensitivity.

\textbf{Insight 3: HPC workloads enable efficient estimation of delay sensitivity.}
There are several challenges associated with estimating delay sensitivity such as:
\begin{enumerate*}[label=(\roman*)]
\item estimating delay sensitivity for a large number of applications, and
\item variability in delay sensitivity across the application lifespan.
\end{enumerate*}
We characterize HPC workloads to show that these challenges do not manifest in (or negligibly impact) production systems and workloads.
First, delay sensitivity needs to be characterized for {\it only} few applications.
\cref{fig:workload_composition} shows the cumulative distribution of node hours utilized by different applications running on three generations of NERSC systems~\cite{nersc_workload}. 
We observe that the top 10 applications comprise nearly 50\% of the workload in these systems.
Second, our characterization of two of these applications reveals traffic stability; thereby, indicating low variability in delay sensitivity across the application lifespan.
\cref{fig:traffic_variation} shows the variation in cumulative traffic injection in two common scientific applications: MILC~\cite{milc} and LAMMPS~\cite{lammps}. 
We observed iterative communication behavior across the run for both applications. 
Moreover, such iterative behaviour has been demonstrated for multiple commonly used HPC applications~\cite{gahvari2011modeling,giannozzi2009quantum,maintz2018strategies}.

\section{\sysname Overview}
\label{s:overview}

\begin{figure}[!t]
    \centering
    \includegraphics{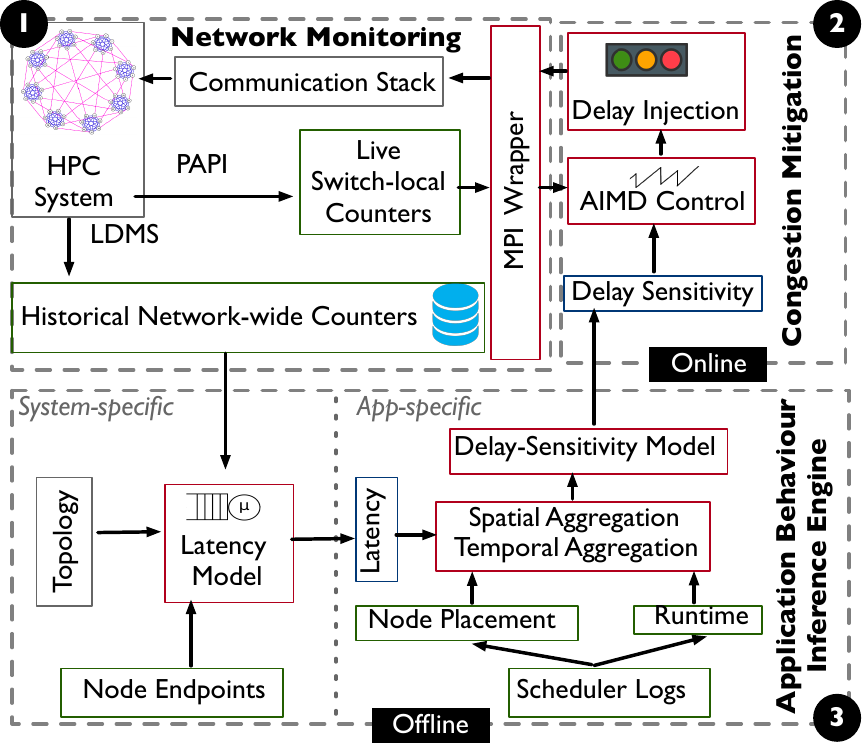}
    \caption{Overview of \sysname}
    \label{fig:approach_overview}
    \spacefigure
\end{figure}

~\cref{fig:approach_overview} depicts the architecture of \sysname, our proposed application-aware congestion mitigation framework for high-performance computing systems.  
The framework consists of two major components: application behaviour inference engine and congestion mitigation driven by existing network monitoring techniques, discussed below.

\noindent \textbf{Network Monitoring (\circled{1}).} 
HPC systems expose network telemetry data of each switch as network performance counters.
\sysname uses network telemetry data to drive
\begin{enumerate*}[label=(\roman*)]
\item the application behavior inference engine, which enables estimation of application delay sensitivity; and 
\item congestion mitigation.
\end{enumerate*}
For the application behavior inference engine, the counters are collected and aggregated across the network by the Lightweight Distributed Metric Service (LDMS)~\cite{agelastos2014lightweight} to provide a coherent global network snapshot.
For the congestion mitigation mechanism, each node periodically samples local switch counters by using the Performance API (PAPI)~\cite{mucci1999papi}.\footnote{We use PAPI because timely mitigation depends on quick access to network performance counters, which is hard to achieve via LDMS, as it aggregates performance counters on the file system.}

Our test HPC systems use the Cray Aries network technology~\cite{alverson2012cray} with a low-diameter Dragonfly topology~\cite{kim2008technology}.
The Dragonfly topology partitions the compute nodes and switches into fully connected ``electrical groups,''~\cite{alverson2012cray} in which each group operates as a high-radix virtual switch.
To mitigate congestion in the Dragonfly network, each node performs static rate control and uses adaptive routing to distribute traffic across network paths~\cite{kim2008technology}.

\noindent \textbf{Application Behavior Inference Engine (\circled{3}).} 
The application behavior inference engine helps us estimate delay sensitivity, a metric that captures the end-to-end effects of congestion. 
The engine is driven by three inter-linked probabilistic regression models:
\begin{enumerate*}[label=(\roman*)]
    \item the \emph{network latency model} (described in~\cref{s:model}), which estimates network latency (i.e., the time required to send a quanta of data\footnote{In Aries, the smallest quantum of data is a flit whose size is $\sim$6 bytes.}) under varying levels of congestion by using network performance counters (indicative of queuing delays), network topology, and compute-node endpoints;
    \item \emph{spatial \& temporal aggregation} (described in~\cref{s:framework}), which estimates expected message delivery time by aggregating inferred network latency spatially (i.e., across multiple communication paths) and temporally (i.e., across the application's lifespan); and
    \item the \emph{delay sensitivity model} (described in~\cref{s:framework}), which estimates application delay sensitivity using inferred message delivery times and corresponding application runtimes.
\end{enumerate*}
The network latency model is system-specific, i.e., it needs to be retrained for each system, whereas the delay sensitivity model and spatial \& temporal aggregation are application-specific, i.e., they need to be retrained for every application configuration.
To train the latency model, we use round-trip time measurements obtained via a specialized \texttt{pingpong} application under diverse congestion conditions and representative compute node endpoints. 
To train the application model, we use application runtimes collected from application runs under synthetically generated or natural system congestion. 

\noindent \textbf{Congestion Mitigation Mechanism (\circled{2}).}
The congestion mitigation mechanism uses a dynamic traffic rate limiter based on an additive increase multiplicative decrease (AIMD)~\cite{chiu1989analysis} control loop whose parameters are inferred using our application behavior inference engine. 
An AIMD control loop exponentially throttles application traffic (multiplicative decrease) when congestion is sensed in the network, and linearly increases traffic (additive increase) in the absence of detected congestion. 
\sysname enhances the state-of-the-art AIMD mechanism\footnote{For example, DCQCN uses AIMD congestion control with a fixed additive increase and multiplicative decrease step for all applications.} by  making the multiplicative decrease step in the AIMD control loop dependent on the application's delay sensitivity, to minimize the penalty (i.e., traffic throttling) on applications that are more sensitive to congestion.

\section{Application Behavior Inference Engine}
This section describes the application behavior inference engine and its components, which include application-specific and system-specific models.
\subsection{Application-specific Models}
\label{s:framework}

Here we present application-specific models that we used to derive the application delay sensitivity, which encapsulates the relationship between network congestion and application runtime as given in \cref{def:dsensitivity}.
To make our methodology of estimating delay sensitivity extensible to a large class of applications, we developed a model that treats an application as a black box and relies solely on the network topology, application node placement, and network telemetry data to infer the delay sensitivity.
\theoremstyle{definition}
    \begin{definition}{\emph{Delay sensitivity}} ($c_a$)
    is the ratio between the increase in application runtime ($T_a$) and increase in the expected value of message delivery times ($M_a$).
    \begin{equation}
        c_a = \frac{\Delta T_a}{\Delta M_a}
        \label{eq:def_dsensitivity}
    \end{equation}
    \label{def:dsensitivity}
    \vspace{-5mm}
\end{definition}

\begin{figure}[!t]
\centering
\includegraphics{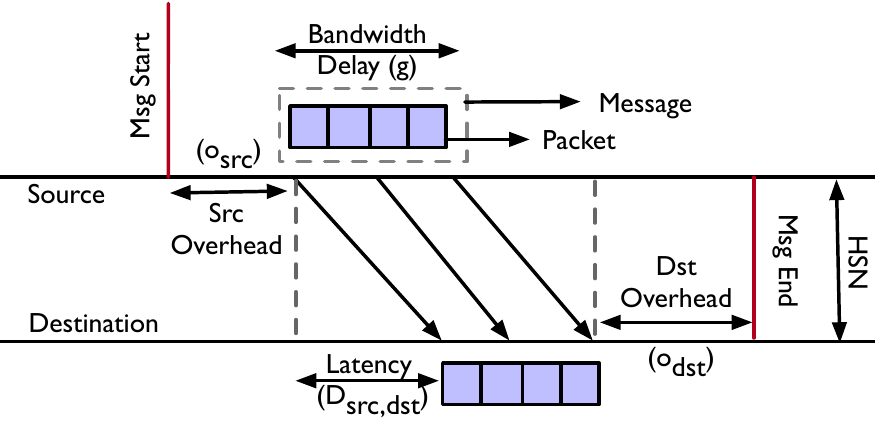}
\caption{Breakdown of total message delivery time}
\label{fig:message_delivery}
\spacefigure
\end{figure}
Our key assumption is that delay sensitivity does not vary across the application's lifespan due to their iterative communication patterns (discussed in ~\cref{s:motivation}). 
As delay sensitivity is constant, the derivative of application runtime with expected message delivery time ($M_a$) is constant, and their interdependence can be modeled by a linear regressor given by~\cref{eq:expected_runtime}. 
\begin{equation}
    T_a = c_aM_a + k_a
    \label{eq:expected_runtime}
\end{equation}
Thus, our quest for estimating delay sensitivity reduces to estimating the slope between expected message delivery times and application runtime. Application runtime can be trivially obtained from scheduler logs. However, estimating expected message delivery times is more involved, and we devote the rest of this section to describing our approach. 

\textbf{Abstracting Message Delivery Time.}
HPC applications typically rely  on sending of messages between two endpoints using a communication library such as the Message Passing Interface. 
The various delays involved in the process of sending/receiving messages are shown in~\cref{fig:message_delivery} (inspired by the logP model~\cite{culler1993logp}). 
A message consists of several packets that are transmitted sequentially across the network from the source to the destination. 
The total message delivery time for a message ($M_{src,dst}(t)$) that is initiated at the source at time $t$ can be broadly categorized into the following components:
\begin{enumerate}[label=(\alph*)]
    \item End-host Overheads ($o_{src}, o_{dst}$): Housekeeping tasks need to be performed at the source (before the packet is sent out) and at the destination (after the packet has been received). These tasks include packet header processing, interface-level processing, and OS scheduling of the packet onto the network interface card (NIC). The total time associated with these tasks are represented by $o_{src}$ and $o_{dst}$ for the source and destination, respectively.
    \item Latency ($D$): The time needed for the first quanta of data to reach the destination is the latency ($D$).
    \item Bandwidth Delay ($g$): The time required to push all the remaining packets onto the network link is the bandwidth delay.
    Given the homogeneity of the network, the bandwidth delays at the source and destination are typically equal. Bandwidth delay is given by the ratio of message size to NIC bandwidth ($\frac{\text{msg size}}{\text{bandwidth}}$).
\end{enumerate}
Therefore, the total message delivery time for a message from source $src$ to destination $dst$ initiated at time $t$ is:
\begin{equation}
    M_{src,dst}(t) = o_{src} + D_{src,dst} + g + o_{dst}
    \label{eq:message_delivery}
\end{equation}

Variations in end-host overheads can be considered noise when the delay introduced by network congestion is the predominant factor in runtime variation. 
Unlike the end-host overheads, bandwidth delay and latency can vary significantly depending on the application and network congestion conditions, which we can estimate through network counters. 
Bandwidth delay can be estimated from the application's message size and injection bandwidth at the network interface card.
However, network latency is not directly observable in the system. Therefore, we infer the network latency metric through performance counters, such as link stall times\footnote{\emph{Stall time} in credit-based flow control networks refers to the time spent waiting to transmit data due to unavailability of egress buffer space.}. Note that $D$ is the one-way latency, which differs from the round-trip time that can be computed directly from the network performance counters. Also, note that one-way latency cannot be computed simply by halving round-trip time measurements because the return path congestion may be significantly different from the forward path congestion. 
If we did not consider reverse path congestion, our model would yield poor estimation of delay sensitivity. 
We will use the network latency model to compute $D$, as we discuss in~\cref{s:model}.

Moreover, because of performance and cost constraints, there are challenges associated with collecting network performance counters for each message individually. 
In particular, on our system, there is a significant  difference between the measurement granularity (1 second) and the message delivery times (on the order of microseconds).
Hence, we apply \cref{eq:message_delivery} for all messages sent in a measurement interval.

The expected message delivery time is dependent on node placement and on variation of congestion over space (spatial characteristics) and time (temporal variation).

\textbf{Accounting for Spatial Characteristics.}
We estimate expected time to deliver a message between any two application nodes by modeling spatial characteristics (i.e., application-to-node mapping and congestion diversity over network links). 
As we have no prior 
information regarding application communication patterns, we assume that any two nodes are equally 
likely to communicate. 
However, when prior knowledge is available, we can deduce which node pairs are more likely to communicate, and use that to improve our estimations.
In practice, we find that without prior knowledge, we can robustly estimate application runtime with high correlation, as shown in \cref{s:results}.
Under the no-prior-knowledge assumption, the expected message delivery time for any node-to-node communication at time $t$ (denoted by $M_s(t)$) for an application with $N$ nodes is given by:%
    \begin{equation}
        M_s(t) = \frac{\sum_{(src,dst)\in N}{M_{src,dst}(t)}}{{{|N|} \choose {2}}}
        \label{eq:spatial_delay}
    \end{equation}
    
\textbf{Accounting for Temporal Variation.}
We model temporal variation (i.e. congestion variation over time) by estimating the expected message delivery time across measurement intervals. The expected message delivery time across the application's lifespan ($M_a(t)$) is estimated using the expected spatial message delivery time ($M_s(t)$), as given by \cref{eq:temporal_delay}:
    \begin{equation}
        M_a = \frac{\sum_{t=0}^{T_a}{M_s(t)}}{T_a},
        \label{eq:temporal_delay}
    \end{equation}
    where $T_a$ is the total application runtime.

\subsection{System-specific Model}
\label{s:model}

In this section, we present a model for estimating the network latency ($D$), i.e., the time required to deliver the first quanta of data after the start of message transmission. We first present the high-level system abstraction, followed by a model for estimating the latency for a single link and its extension to a set of paths.

\noindent \textbf{System Abstraction.}
In our system abstraction, we represent each hop between the source and destination nodes as a link. 
In a high-speed interconnect, examples of such links are connections between the processor, NIC, and network tiles/ports on intermediary switches. 
The interconnect is lossless and uses a credit-based hop-by-hop flow control mechanism~\cite{credit1995kung}. 
To transmit data, the sender requires credits from the receiver, which are 
calculated based on queue occupancy. 
When there are insufficient credits available to transfer the waiting data, 
the data transmission stalls, and the data will be buffered in the link buffer until sufficient credits become available. 
The process of 
waiting in the queue can cause head-of-line blocking, which can lead to cascading backpressure in 
adjacent links and formation of congestion trees across the interconnect~\cite{escudero2014efficient}.

The time quanta each link waits to send data is referred to as a \emph{stall}. Network counters record the 
total time each link is stalled ($T_{s}$). Using these link stall times, we compute the derived metric for each link, Percent Time Stalled ($P_{Ts}$).
\begin{equation}
    P_{T_s} = 100\times\frac{T_{s}}{T_{m}},
\end{equation}
where $T_s$ is the total stall duration during the measurement interval and $T_m$ is the measurement interval.

\noindent \textbf{Single-Link Latency Estimation.}
To estimate the latency for a single link, we break it down into the following components:
\begin{enumerate}[label=(\alph*)]
\item \emph{Processing delay} is the time the switch needs to process packet headers and move the packet along the queue before further routing.
\item \emph{Transmission delay} is the time needed to push all of the packet’s bits onto the wire.
\item \emph{Queuing delay} is the time that the data to be transmitted spends in the link buffers awaiting transmission.
\item \emph{Propagation delay} is the time needed for data to transit the wire, which depends on the number of hops needed to route the message from `src' to `dst' and wire length.
\end{enumerate}

Both the processing and the propagation delays are dependent on the transmission media latency and the packet header processing overhead. 
When modeling a single link, we assume that these delays are constant. 
Queuing delays dominate over transmission delays because we consider only a small quanta of data and the time required to transmit the quanta into the link is negligible compared to queuing delays.

For the sake of modeling, we represent whether the link buffers are serviced or freed by a random probabilistic process. At each time unit, the link state can be either stalled or non-stalled. During the non-stalled state, the buffer is emptied and the data are allowed to be transmitted. Thus, the link state at each time unit can be represented by a Bernoulli random variable with parameter $p$ that takes value 1 if the link is in a stalled state and 0 otherwise. To infer the total waiting time, we need to compute the expected number of consecutive time units for which the packet waits in the link buffer. An appropriate discrete distribution to model the above-mentioned constraints is a Geometric distribution.
\begin{equation}
    P[T = t] = (1-p)p^{t-1}
\end{equation}

The expected waiting time is given by Equation (\ref{eq:expected_time}). 
\begin{equation}
    E[T] = \sum_{t=1}^{\infty}{t(1-p)p^{t-1}} = \frac{1}{1-p}
    \label{eq:expected_time}
\end{equation}

To estimate the parameter $p$, we use the maximum likelihood estimation (MLE)~\cite{mle}. The log-likelihood of measuring $T_s$ units of stall in the measurement interval $T_m$ is given by Equation (\ref{eq:likelihood}). 
\begin{align}
\label{eq:likelihood}
log(\mathcal{L}) & = log(p^{T_s}(1-p)^{T_m-T_s}) \\ \nonumber
                 & = T_slog(p) + (T_m-T_s)log(1-p)
\end{align}

The optimal parameter value can be obtained by setting the gradient of $log(\mathcal{L})$ to zero. Then the value of $p$ that maximizes the log-likelihood of observing the given stalls is given by Equation (\ref{eq:max_p}).
\begin{equation}
    p = \frac{T_s}{T_m} = \frac{P_{Ts}}{100}
    \label{eq:max_p}
\end{equation}

The expected queuing delay for the link ($Q_l$) is given by:
\begin{equation}
    Q_l = E[T] = \frac{100}{100-P_{Ts}}
\end{equation}

\noindent \textbf{Extension to Multiple Paths.}
To estimate the network latency using link-level queuing delays ($Q_l$), we use a linear regression model (refer to \cref{eq:network_latency}. 
Let the possible paths (i.e., a set of links) between two nodes $n_1$ and $n_2$ be represented by the sets $S^*$. Each set $S_i$ that belongs to $S^*$ is composed of a set of links that make up a unique path between the two nodes.
Note that this path comprises all the intermediary links that include connections between processors, NICs, and network tiles on switches. 
\begin{equation}
    D_{n_1,n_2} = \sum_{S_i \in S^{*}}k_s\sum_{l \in S_i} k_{i_1}Q_l+ k_{i_2},
\label{eq:network_latency}
\end{equation}
where $S^*$ is the set of all links on all possible paths as constrained by the routing algorithm~\cite{kim2008technology} between 2 nodes $n_1$ and $n_2$. The constants $k_i$ are estimated using training. %
\section{Mitigation}
\label{s:mitigation}
\sysname's online congestion mitigation consists of the following major components:
\begin{enumerate}[label=(\alph*)]
    \item The \emph{congestion signal}, which is an estimation of the congestion experienced by packets. It is calculated from network performance counters sampled at the NIC.
    \item The \emph{delay injection module}, which provides a mechanism for controlling congestion at the source by introducing probabilistic delays between consecutive message send requests.
    \item \emph{Rate control}, which uses the delay injection module to dynamically change the traffic injection rate based upon the congestion signal.
\end{enumerate}

\textbf{Congestion Signal.}
We choose to use stalls experienced when injecting traffic from the NIC into the high-speed network (HSN) as the signal for congestion. 
While network latency (described in \cref{s:model}) would be an ideal congestion signal (as it can estimate the impact of congestion on application runtime), such a metric is difficult to calculate during online mitigation as we have access to network performance counters on local switches \textit{only}. 
In lieu of network latency, we use stalls experienced by the application traffic injected from the NIC to the HSN (referred to as $nic2hsn$) as it captures both local congestion (i.e., backpressure applied by the NIC) and outside congestion (i.e., backpressure imposed by the network).
We collected the $nic2hsn$ counters by using the PAPI~\cite{mucci1999papi} interface with the sampling frequency set at one second.

\textbf{Delay Injection Module.}
As no direct traffic rate-throttling method is available on the compute nodes, we introduce delays between consecutive message requests sent out by every process. 
The delay is the time delta between traffic injection at the line rate and traffic injection at the desired reduced rate:
\begin{equation}
    \Delta T_{inj} = \frac{R_{line}-R_{req}}{B_{nic}},
    \label{eq:delay_injection}
\end{equation}
where $\Delta T_{inj}$ is the delay injected between messages, $R_{line}$ is the line rate at which messages are sent out, $R_{req}$ is the required reduced rate, and $B_{nic}$ is the NIC bandwidth.

It is possible that $\Delta T_{inj}$ may be on the order of nanoseconds; however, Linux only supports process sleep with microsecond resolution. 
To overcome this challenge, we use probabilistic delays. For example, if the delay that must be introduced is 0.3$\mu$s, we introduce a 1$\mu$s delay with a probability of 0.3. 
The probability is simulated using a uniform random number generator from the Linux standard library. We achieve the following using an MPI wrapper as shown in \cref{fig:approach_overview}.

\textbf{Rate Control.}
While there are a plethora of rate-control algorithms based on Additive Increase Multiplicative Decrease (AIMD) rate control~\cite{chiu1989analysis}, we decided to adapt the rate-control algorithm first proposed in DCTCP~\cite{alizadeh2010data} for general-purpose datacenters and later used in DCQCN~\cite{zhu2015congestion} for RDMA over converged Ethernet (RoCE) networks. 
We selected the DCQCN rate control algorithm because it has been widely adopted in production networks because of its inherent simplicity and consequent ease of deployment.
The sender maintains an estimate of stalled traffic called $\alpha$, which is updated every measurement window using the following update:
\begin{equation}
    \alpha \leftarrow (1-g)\alpha + g \cdot nic2hsn,
\end{equation}
where $g \in (0,1)$ is the weight given to past samples as opposed to past measurements. Consequently, the traffic rate injected by application $a$ at time $t$ (i.e., $R_a$) is updated as follows:

\begin{equation}\label{eq:r_update}
R_a \leftarrow
\begin{cases} 
    R_a(1-\alpha/(2+c_a)),& \text{if } \alpha>0\\
    R_a+1,                & \text{otherwise,}
   \end{cases}
\end{equation}
where $c_a \geq 0$ is the delay sensitivity of the application $a$.
Thus, $\alpha$ and $c_a$ together control the sending rate. A higher value of $\alpha$ (high congestion) throttles the traffic rate more aggressively, while a higher value of $c_a$ (more sensitive to network congestion) prevents the traffic rate from being reduced more than necessary. Note that $c_a$ is a term introduced by us in the multiplicative decrease step.
We leverage and extend the existing fairness and convergence properties of the AIMD rate controller to ensure that the modified AIMD controller also converges and is fair. We omit the proofs here because of space constraints.

\section{Evaluation}
\label{s:results}

In this section, we evaluate the effectiveness of our application behavior inference engine and demonstrate the performance gain obtained using our application-aware congestion mitigation. 
We performed the evaluation across four Cray Aries~\cite{alverson2012cray} HPC systems of varying scales: 
\begin{enumerate}[label=(\alph*)]
    \item \emph{Voltrino} consists of a chassis with 16 switches and 240 network links.
    \item \emph{Mutrino} consists of an electrical group\footnote{384 compute nodes and 96 switches are combined to create an electrical group that acts as a single high-radix switch.} with 96 switches and 2,880 network links.
    \item \BLUE{\emph{Neutrino} consists of two electrical groups with 192 switches and 6,720 network links.}
    \item \emph{Cori} is a production-scale supercomputer located at \emph{NERSC} comprising 34 electrical groups and 130,560 network links. 
\end{enumerate}

We used these systems to evaluate our models in testbed and production settings. 
For example, Voltrino and Mutrino are testbed systems that we used to conduct  multiple application runs and rigorously validate our models. 
We used to Cori to validate our methodology in production settings.
\BLUE{Neutrino is a customized HPE/Cray system with an over-provisioned network (i.e., it has higher inter-group connectivity than a typical Aries system). 
Neutrino is not available to users; however, HPE/Cray was able to provide us with benchmark traces designed to understand network congestion.}

Our application set consists of both representative benchmarks (a--c) and real applications (d--g), as described below:
\begin{enumerate}[label=(\alph*)]

\item\BLUE{\emph{DSA} is a synthetic delay sensitive application that implements a sorting algorithm.}

\item\BLUE{\emph{BIA} is a synthetic bandwidth intensive application that uses the Cray SHMEM library to stress the network.}

\item \emph{GPCNeT} is a congestion-creation benchmark suite that captures four congestion patterns (all-to-all, RMA\footnote{\emph{RMA} stands for \emph{remote memory access}.} incast, point-to-point incast, and RMA broadcast) commonly found in HPC workloads~\cite{chunduri2019gpcnet}.

\item \emph{MILC} is used to study quantum chromodynamics, the theory of strong interactions of subatomic physics. It performs four-dimensional lattice communication, and its performance limiters include network latency~\cite{milc}.

\item \emph{LAMMPS} is a molecular dynamics program focused on materials modeling~\cite{lammps}. Performance limiters include compute, memory bandwidth, network bandwidth, and network latency~\cite{coral_2}.

\item \emph{Quantum Espresso} is an integrated application suite for electronic-structure calculations and materials modeling~\cite{giannozzi2009quantum}. Performance limiters include compute and network latency.~\cite{quantum_benchmarking}

\item \emph{AMG} is a parallel algebraic multigrid solver for linear systems~\cite{amg}. Performance limiters include memory-access and network latency~\cite{coral_2}.
\end{enumerate}

\BLUE{Applications (a--b) were provided to us by HPE/Cray to help us understand network congestion mitigation capabilities.} Application c is an open-source congestion benchmark suite. We chose applications (d--g) because
\begin{enumerate*}[label=(\roman*)]
\item they are among the top 10 most commonly used applications on NERSC production systems\footnote{MILC and Quantum Espresso are the top two open-source applications on NERSC systems, accounting for 12\% of total node hours.}~\cite{nersc_workload}, and 
\item previous work has shown that these applications are very susceptible to congestion-induced performance degradation~\cite{bhatele2020case}.
\end{enumerate*}

\subsection{Evaluating the Application Behavior Inference Engine}
\begin{figure*}[ht]
    \centering
    \begin{subfigure}[c]{0.325\textwidth}
    \includegraphics{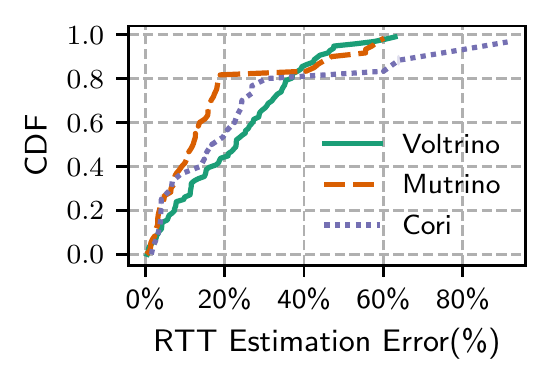}
    \caption{Distribution of percentage error in RTT estimation for Voltrino, Mutrino, and Cori}
    \label{fig:latency_error}
    \end{subfigure}
	\begin{subfigure}[c]{0.325\textwidth}
    \includegraphics{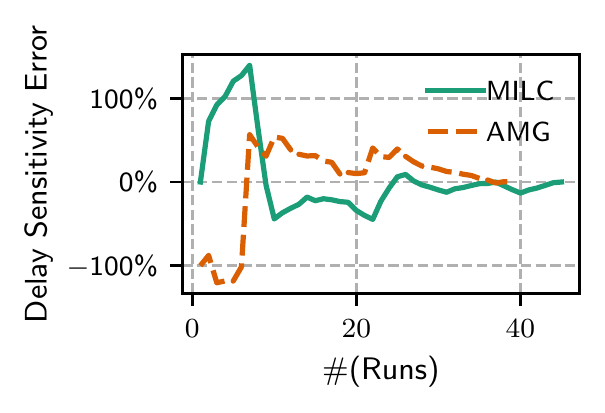}
    \caption{Delay sensitivity convergence with increase in production application runs}
    \label{fig:delay_sensitivity_convergence}
    \end{subfigure}
    \begin{subfigure}[c]{0.325\textwidth}
    \includegraphics{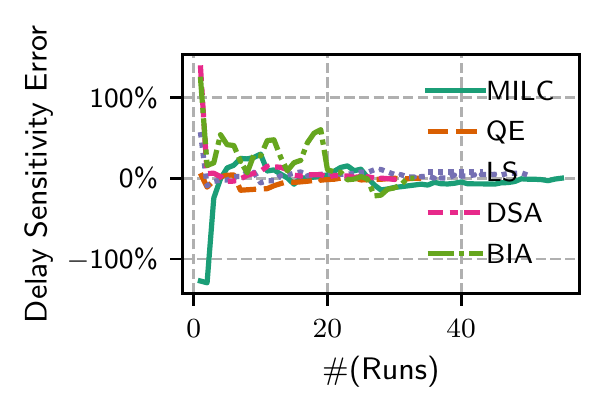}
    \caption{Delay sensitivity convergence with increase in synthetically congested application runs}
    \label{fig:delay_sensitivity_convergence_controlled}
    \end{subfigure}
\caption{Evaluating convergence and accuracy of application behavior inference engine}
\end{figure*}
\textbf{Training the Network Latency Model.}
The network latency model uses round-trip time (RTT) measurements recorded using a \texttt{pingpong} application to generate training labels, and network performance counters and node endpoints as inputs.
We use RTT measurements because it is challenging to estimate one-way network latency directly in the absence of perfect time synchronization (which is infeasible in practice).
We executed the \texttt{pingpong} application multiple times (approximately forty) on two randomly chosen nodes in the presence of congestion created using GPCNeT.
During the \texttt{pingpong} application training runs, the percent time stalled measurements (obtained using network counters) across all links in the system ranged between 0\% and $99$\%.
Each \texttt{pingpong} run involved 10,000 consecutive message exchanges (without any compute), and the average RTT for each message exchange was recorded. 
A large number of messages were sent to compensate for the difference in network latency (2--100 $\mu$s) and granularity of measurements (1 s). 
To train the model, we sum the forward and reverse path latency that are calculated using Equation \cref{eq:network_latency} to estimate the round-trip time. Thus, we can calculate the constants ($k_i$) using the least sum of squares fit with training data (i.e. RTT measurements obtained from the \texttt{pingpong} application).

\textbf{Accuracy of Latency Model.}
We evaluate the accuracy of the latency model, described in~\cref{s:model}, which is used to estimate network latency by using performance counters. The model was trained and validated on \emph{Voltrino}, \emph{Mutrino}, and \emph{Cori}.
\cref{fig:latency_error} shows the percentage error in estimating RTT for the three systems. For each of the three systems, 75\% of the RTT measurements were estimated within an error bound of 30\%. 
Moreover, we observed statistically high Pearson correlation coefficients~\cite{Benesty2009} of 0.68, 0.71, and 0.67 between the measured and the estimated round-trip time for Voltrino, Mutrino, and Cori, respectively.
\emph{The latency model remains accurate under large variation in congestion and system scale.}

\begin{table}[]
\centering
\resizebox{\columnwidth}{!}{%
\begin{tabular}{@{}ccccc@{}}
\toprule
Application & Nodes & System & \begin{tabular}[c]{@{}c@{}}Congestion\\ Source\end{tabular} & \begin{tabular}[c]{@{}c@{}}Placement\\ Decision\end{tabular} \\ \midrule
DSA     & 128 & Neutrino     &  BIA   & Random           \\
 BIA   & 64  & Neutrino     &  BIA   & Random           \\
MILC-small        & 4   & Mutrino & GPCNeT            & Random          \\
Quantum  Espresso & 4   & Mutrino & GPCNeT            & Random          \\
LAMMPS            & 4   & Mutrino & GPCNeT            & Random          \\
MILC              & 384 & Cori    & System Congestion & Slurm Scheduler \\
AMG               & 768 & Cori    & System Congestion & Slurm Scheduler \\
 \bottomrule
\end{tabular}%
}
\caption{List of applications (and their configurations) used to evaluate the application behavior inference engine}
\label{tab:app_engine_in}
\end{table}

\begin{table}[]
\centering
{%
\begin{tabular}{@{}ccccc@{}}
\toprule
Application &
  \begin{tabular}[c]{@{}c@{}}Mean Runtime\\  (sec)\end{tabular} &
  \begin{tabular}[c]{@{}c@{}}Runtime Fit\\ Correlation\end{tabular} &
  Delay Sensitivity \\ \midrule
  DSA     & 448 & 0.90 & 1.3  \\
  BIA   & 242 & 0.73 & 0.18 \\
MILC-small        & 220 & 0.71 & 6.67 \\
Quantum  Espresso & 250 & 0.89 & 12.3 \\
LAMMPS            & 23  & 0.81 & 0.47 \\
MILC              & 848 & 0.73 & 45   \\
AMG               & 459 & 0.70 & 12.5   \\ \bottomrule
\end{tabular}%
}
\caption{Output generated by the application behavior inference engine for different applications}
\label{tab:app_engine_out}
\spacefigure
\end{table}

\textbf{Accuracy of Estimating Delay Sensitivity.}
We have designed a comprehensive test suite for validating the delay sensitivity ($c_a$) estimator (introduced in Section~\ref{s:framework}) that uses the configurations described in Table~\ref{tab:app_engine_in}. 
For smaller, single-user testbed systems (i.e., \BLUE{Mutrino and Neutrino}), we use synthetically generated congestion to create runtime variation.
However, on larger, multi-user systems like Cori, we find sufficient natural runtime variation due to network contention between different applications.

To test whether the delay sensitivity estimator is sensitive to application node placement, we randomly varied the application-to-node mapping for the above-mentioned applications on the smaller-scale systems. 
For Cori, the node placement was determined by the Slurm scheduling policy~\cite{yoo2003slurm}, and we used a one-year production dataset, as it provided sufficient diversity in both placement and system congestion.
\BLUE{Moreover, our test suite consists of both highly delay sensitive (e.g., Quantum Espresso) and bandwidth intensive (e.g., BIA) applications.}

Table~\ref{tab:app_engine_out} provides the delay sensitivity values for our test suite. 
We found that message delivery times have a strong correlation between 0.7 and 0.9 for all applications in the test suite. 
\BLUE{The estimated delay sensitivity could vary between zero (implying negligible increase in application runtime due to congestion; e.g., BIA)} to a very high value (implying significant increase in application runtime due to congestion; e.g., MILC).
The delay sensitivity is calculated with respect to the \texttt{pingpong} application, i.e., the \texttt{pingpong} application has a delay sensitivity equal to one.
The proposed delay sensitivity metric is robust as it captures the relationship between application size and congestion.
In our experiments, we observed that increasing the application size increased the delay sensitivity, thereby confirming our intuition that applications with larger size have larger number of communication paths, and thus link delays are more likely to stall the entire iteration.
For example, delay sensitivity increases by a factor of 6.7\texttimes~ for MILC when the number of nodes increases from 4 to 384.
In addition, as expected, we observed that applications that are only latency-bound have higher delay sensitivity than applications that have multiple performance limiters, such as compute and memory. For example, Quantum Espresso, which is solely latency-bound, has a higher delay sensitivity than LAMMPS, which is also limited by memory and compute.
\emph{We find that the delay sensitivity estimator is accurate for a wide range of applications, scales, congestion and placements.} 

\textbf{Training Requirements for Delay Sensitivity Estimation.}
We quantify the cost of training our models to accurately estimate the delay sensitivity ($\hat{c}_a$)in terms of the number of application runs.
\cref{fig:delay_sensitivity_convergence} shows the error in estimating the delay sensitivity with increasing number of production training runs for MILC and AMG. The error in delay sensitivity at the $i^{\text{th}}$ run is given by $\frac{c_a^{i}-\hat{c}_a}{\hat{c}_a} \times 100$. For both applications, we observe that the delay sensitivity fluctuates significantly at the beginning of the training, but stabilizes to within 10\% of the final estimated delay sensitivity ($\hat{c}_a$) after $\sim$25 runs.
However, we can further reduce the training time significantly by exposing the application to both extremes of congestion (i.e., no congestion and high congestion),
thus providing early benefits to the user.
As we show in ~\cref{fig:delay_sensitivity_convergence_controlled}, the training time is reduced by 5\texttimes~ (from 25 runs to 5 runs) in the case of MILC running under synthetic congestion. We observed similar trends for Quantum Espresso (labeled ``QE'' in the figure), LAMMPS (labeled ``LS''), \BLUE{DSA (labeled ``DSA''), and BIA (labeled ``BIA'').} 
\emph{Our model converges fast, especially under synthetic congestion.}

\subsection{Evaluating Congestion Mitigation}
We compared \sysname's application-aware congestion mitigation mechanism (described in~\cref{s:mitigation}) to DCQCN rate control (deployed in RDMA networks) and the static rate control mechanism in Cray. 
In these experiments, we considered two workloads: workload W1 that consists of MILC and GPCNeT, and workload W2 that consists of two GPCNeT instances. These workloads were run on Voltrino.
W1 and W2 were chosen to represent two extreme ends, i.e., W1 tests a combination of a delay sensitive application and a bandwidth intensive application, and W2 tests a combination of two bandwidth intensive applications. 
We do not consider a workload consisting of two delay sensitive applications, as they would be unable to create sufficient congestion.
Moreover, we {\it only} consider two applications at a time because mitigation is performed individually for each application, and is dependent only on network congestion (created by the workload) and not on the number of applications. 
Each application was randomly placed in the system to generate various levels of congestion across the network links.
Overall, we executed 334 runs to compare each workload with every congestion mitigation mechanism separately.

\begin{figure*}[ht]
    \centering
	\begin{subfigure}[c]{0.325\textwidth}
    \includegraphics{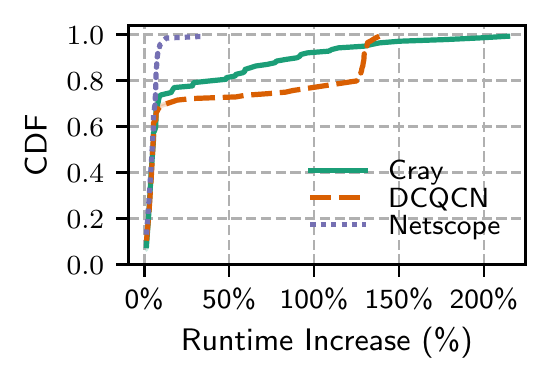}
    \caption{Comparing runtime variability across Cray, DCQCN, and \sysname}
    \label{fig:runtime_increase}
    \end{subfigure}
    \begin{subfigure}[c]{0.325\textwidth}
    \includegraphics{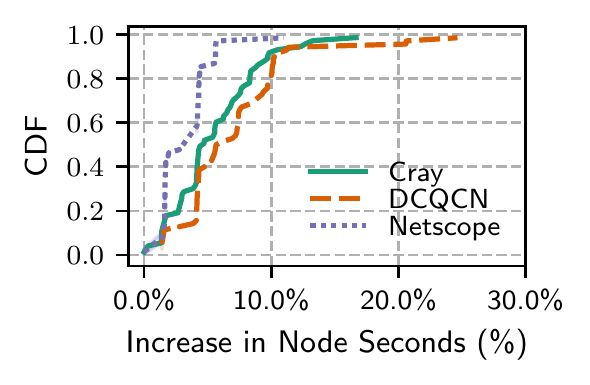}
    \caption{Comparing total compute requirements (node seconds required for completion) across Cray, DCQCN, and \sysname}
    \label{fig:node_seconds}
    \end{subfigure}
    \begin{subfigure}[c]{0.325\textwidth}
    \includegraphics{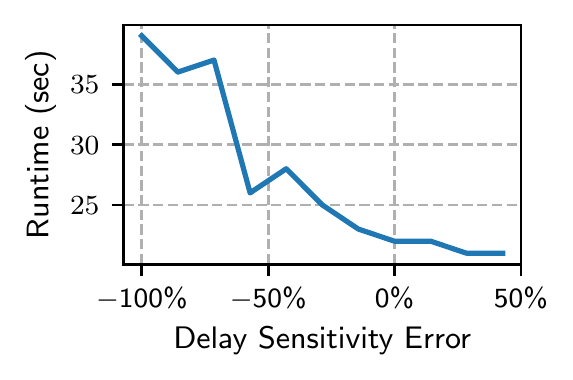}
    \caption{Impact of error in estimating delay sensitivity on MILC runtime when running workload W1}
    \label{fig:training_error}
    \end{subfigure}
\caption{Comparing system throughput of different congestion control mechanisms}
\label{fig:prediction}
\spacefigure
\end{figure*}

\textbf{Reduced Runtime Variability.}
~\cref{fig:runtime_increase} shows the CDF of the percentage runtime increase for two applications (MILC and GPCNeT) across all experiments. 
We found that \sysname outperforms both Cray and DCQCN in reducing the overall runtime variability. 
In particular, at the 99th percentile, \sysname is 11.2\texttimes~better than DCQCN and 14.9\texttimes~better than Cray in reducing the percentage runtime increase.
\sysname is significantly better than DCQCN and Cray in the following respects.
\begin{enumerate*}[label=(\roman*)]
\item Compared to DCQCN, \sysname penalizes the delay sensitive application less aggressively than GPCNeT as the delay sensitive application has a smaller multiplicative decrease factor (described in \cref{s:mitigation}).  
\item Compared to Cray static rate control, \sysname prevents extreme congestion cases because it dynamically adjusts traffic rates based on measured congestion. 
\end{enumerate*}

 \textbf{Increased System Throughput.}
We measured overall system throughput in terms of the node seconds required to complete the workload.
~\cref{fig:node_seconds} shows the CDF of the percentage increase in total node seconds required to run the workloads using the three congestion mitigation mechanisms.
In particular, at the 99th percentile, \sysname is 2.8\texttimes~ better than DCQCN and 1.9\texttimes~ better than Cray in terms of system throughput.
Moreover, we found that DCQCN's system throughput is worse than that of Cray static rate control.
The lower system throughput for DCQCN also explains why AIMD-based algorithms, which are widely used in both wide-area networks and cloud datacenters, have not been widely used in HPC systems.

\begin{table}[]
\centering
{%
\begin{tabular}{@{}cccll@{}}
\toprule
         & \multicolumn{2}{c}{Latency ($\mu$s)} & \multicolumn{2}{l}{Traffic (Flits/sec)} \\ \midrule
         & Median         & 90\%ile         & Median             & 90\%ile            \\
Cray     & 10.1           & 52              & 2.9e7              & 5.4e7              \\
DCQCN    & 3.4            & 11.9            & 2.1e7              & 5.1e7              \\
\sysname & 9.7            & 12.2            & 4.2e7              & 5.2e7              \\ \bottomrule
\end{tabular}%
}
\caption{Comparing traffic and latency for Cray, \sysname, and DCQCN.}
\label{tab:network_level}
\spacefigure
\end{table}

\textbf{Impact on Traffic and Congestion.}
Table~\ref{tab:network_level} shows the network latency (i.e., time required to deliver a flit) and traffic for the three rate-control algorithms across all experiments.
We found that DCQCN has a lower network latency than either Cray or \sysname (by 2.9\texttimes) at the 50th percentile.
Thus, DCQCN is significantly better at reducing network congestion than either \sysname or Cray.
However, solely attempting to lower congestion (i.e., minimizing queuing at link buffers) may adversely affect delay sensitive applications, leading to worse system throughput as discussed above.
Unsurprisingly, static rate control, which does not adapt to varying levels of congestion, performs worse at the 90th percentile, as seen for the Cray rate-control mechanism.
\emph{Overall, we found that \sysname outperforms both DCQCN and the Cray static rate control mechanism at reducing application runtime variation and improving system throughout. }

\textbf{Impact of Training Errors on Congestion Mitigation.}
Here we evaluate the impact of inaccurately estimating delay sensitivity on congestion mitigation efficiency.
Such inaccurate estimations occur because of insufficient training.
A negative error in estimated delay sensitivity, i.e., causing the estimated value to be lower than the true value, may increase application runtime. 
The reason is that a delay sensitive application would have a more aggressive multiplicative decrease rate control step, leading to more severe penalization. 
Further, we cannot afford to have a positive error in estimating delay sensitivity, as it is preferable to have a stronger multiplicative decrease step to mitigate congestion quickly.
~\cref{fig:training_error} shows the runtime for MILC (in workload W1) with error introduced in the delay sensitivity estimation. 
We found that the runtime is close to minimal when the estimation error is close to 0\%. As the negative error increases, the MILC runtime worsens.
In particular, at an error of -100\% in estimating delay sensitivity, \sysname's performance is no better than DCQCN (which is the current state of the art).
\emph{If the estimated delay sensitivity is lower than the actual delay sensitivity, \sysname's performance is worse. 
However, at the same time, we want the estimated delay sensitivity to be as low as possible, to ensure effective congestion control.}

\noindent \textbf{Testing \sysname in Production Systems.}
We tested \sysname on Cori with a scaled-up workload W1, i.e., MILC (baseline runtime 149 s) and GPCNeT (baseline runtime 75 s) were scaled up to 108 and 100 nodes, respectively.
This modified workload occupied $\sim$75\% of an electrical group to minimize the effect of congestion due to other applications, thereby allowing us to conduct controlled experiments. 
We compared the performance of \sysname with that of the Cray and DCQCN rate-control mechanisms. 
For all three rate-control mechanisms, GPCNeT had minimal variation of a few seconds. 
With \sysname, the MILC runtime was restricted to 153 s (1.02\texttimes~worse). On the other hand, DCQCN resulted in a runtime increase to 180 s (1.2\texttimes~worse), and Cray static rate control resulted in a runtime increase to 200 s (1.3\texttimes~worse). 
Note that the production environment introduces additional constraints such as:
\begin{enumerate*}[label=(\roman*)] 
\item Slurm uses a best-fit algorithm based on number of consecutive nodes~\cite{yoo2003slurm}. 
In our case, the allocation policy enforces several communication paths to be within the same switch; thereby, limiting the overall congestion experienced by the applications\footnote{In real-world production runs, jobs span several groups and switches resulting in significant runtime variation and effects of congestion~\cite{bhatele2020case}.}; and
\item the unavailability of large number of node-hours to execute system experiments and characterize tail behavior. 
For example, we {\it only} executed six experiments in total to demonstrate the efficacy of \sysname in large-scale production systems.  
\end{enumerate*}
\emph{Despite the constraints introduced in production, we find that performance benefits of \sysname continue to hold with larger application sizes\footnote{We ran MILC and GPCNeT on 4 nodes and 18 nodes on Voltrino. In contrast, we scaled MILC and GPCNeT by 27\texttimes~and 5.6\texttimes~scaling on Cori.}.} \\
\textbf{Overhead of \sysname.}
Although \sysname increases the congestion-free runtime for the tested applications by up to 2\%, the benefits of \sysname are realized in the presence of congestion. Without \sysname, congestion could cause runtime variation as high as 200\% (see \cref{fig:runtime_increase}).
\sysname's overhead is negligible in terms of memory ($\sim$0.8KB) and computation. %
\section{Related Work}
\label{s:related}

\noindent \textbf{Modeling impact of congestion on HPC applications.}
Several researchers have studied the effects of network congestion on the performance of HPC applications. ~\cite{bhatele2020case} has proposed blackbox machine learning models to estimate application runtime from network counters. ~\cite{jha2020measuring} has proposed a congestion region-based approach to characterize congestion in torus networks. However, these approaches do not provide application characteristics that can be consumed by the CC algorithm.

\noindent \textbf{Rate Control for Congestion Mitigation.}
Traditionally, Infiniband congestion mitigation~\cite{gran2010first} has been the defacto standard for CC in credit-based flow control networks. In addition, researchers have proposed additional schemes~\cite{escudero2014efficient, luo2012congestion,jiang2015network} to mitigate congestion for HPC systems. Moreover, there are a large number of application-oblivious(e.g., \cite{alizadeh2010data,mittal2015timely,zhu2015congestion}) and application-aware (e.g., \cite{D2TCP, PDQ,grosvenor2015queues}) CC mechanisms proposed for cloud datacenters. However, the application-oblivious CC mechanisms end up incorrectly penalizing applications whereas the application-aware CC mechanisms require complicated parameter tuning that is difficult to achieve without an automated framework. Cray Slingshot~\cite{de2020depth} proposes to alleviate congestion by selectively targeting the true sources of congestion.

\noindent \textbf{In-network Mechanisms.}
The proposed delay sensitivity metric can be used to assign priority classes to provide quality-of-service.
For example, we can use the delay sensitivity metric to assign priority classes in Cray Slingshot~\cite{de2020depth}. 
In addition, there has been a large body of work on using different packet scheduling policies~\cite{PIFO,pFabric,LSTF} at the switches to minimize impact of congestion on applications that do not contribute to congestion. However, these policies have yet to be deployed on a real switch hardware. %
\section{Conclusion and Future Work}
\label{s:conclusion}
We propose \sysname, an ML-driven framework that achieves dynamic congestion mitigation based on application characteristics and network telemetry.
We comprehensively evaluate \sysname on multiple systems and applications.
\sysname is shown to reduce application tail runtime variability on a Cray Aries testbed system by 14.9\texttimes~while increasing median system utility by 12\%.

There are several open questions and research challenges regarding \sysname that we plan to address in the future: \begin{enumerate*}[label=(\roman*)]
    \item using a network simulator, such as traceR~\cite{jain2016evaluating}, to train the ML models instead of using real application runs to reduce the training costs further;
    \item understanding the impact of variation in application input deck on delay sensitivity; and 
    \item using delay sensitivity to drive in-network mechanisms such as QoS classes and programmable packet scheduling.
\end{enumerate*}

{
    \Urlmuskip=0mu plus 1mu\relax
	\bibliographystyle{IEEEtran}
	\bibliography{bibliography.bib}
}
\end{document}